\documentclass{PoS}

\usepackage{xspace,amsmath}

\newcommand{\eftnopi}{EFT(${\pi\hskip-0.55em /}$)\xspace}
 
\newcommand{\wave}[3]{\ensuremath{{}^{#1}\mathrm{#2}_{#3}}\xspace}
\newcommand{\oneS}{\wave{1}{S}{0}}

\newcommand{\threeS}{\wave{3}{S}{1}}

\newcommand{\VS}{\vec{\sigma}}

\newcommand{\LRd}{\overset{\leftrightarrow}{D}}

\newcommand{\gA}{g^{(^3 \! S_1-^1 \! P_1)}}
\newcommand{\gB}{g^{(^1 \! S_0-^3 \! P_0)}_{(\Delta I=0)}}
\newcommand{\gC}{g^{(^1 \! S_0-^3 \! P_0)}_{(\Delta I=1)}}
\newcommand{\gD}{g^{(^1 \! S_0-^3 \! P_0)}_{(\Delta I=2)}}
\newcommand{\gE}{g^{(^3 \! S_1-^3 \! P_1)}}

\title{Hadronic parity violation in effective field theory}

\ShortTitle{Hadronic parity violation}

\author{\speaker{M.~R.~Schindler}\\
        Department of Physics and Astronomy\\
        University of South Carolina\\
        Columbia, SC 29208\\ USA\\
        E-mail: \email{schindler@sc.edu}}

\abstract{The application of effective field theories to hadronic parity violation in two- and three-nucleon systems is described.
These methods provide several important advantages over the traditionally used meson-exchange models, e.g., model independence and the possibility to reliably estimate theoretical errors.
We focus on the so-called ``pionless'' theory and describe calculations of various two- and three-nucleon observables. The role of parity-violating three-nucleon interactions is also discussed.}

\FullConference{The 7th International Workshop on Chiral Dynamics,\\
		August 6 -10, 2012\\
		Jefferson Lab, Newport News, Virginia, USA}

\begin{document}

\section{Introduction}

The weak interaction between quarks induces a parity-violating (PV) component in nucleon interactions (for reviews, see, e.g., Refs.~\cite{Adelberger:1985ik,Haeberli:1995uz,RamseyMusolf:2006dz}).
Compared to the parity-conserving (PC) parts of the interaction, this component is typically suppressed by factors of $10^{-6}$ to $10^{-7}$.
To measure such small effects, one considers pseudoscalar observables which would vanish if parity was conserved. 
These observables typically are proportional to the correlation between a spin and a momentum ($\sim\vec{\sigma}\cdot \vec{p}$) of the particles involved in the reaction of interest.
Examples include longitudinal and angular asymmetries, as well as induced polarizations. 
Traditionally, the PV interaction between two nucleons has been described in terms of meson-exchange models, in which the exchanged meson couples to the nucleons via one PC and one PV vertex. 
While the PC vertices are typically well-known, the PV vertices are much more difficult to determine.
Reference \cite{Desplanques:1979hn} (referred to as DDH in the following) used a meson-exchange model with $\pi$, $\rho$, and $\omega$ mesons and tried to determine ``reasonable ranges'' for a number of PV meson-nucleon couplings based on quark models and symmetries arguments. 
The DDH model and associated couplings have been the standard for analyzing and interpreting experiments for the past decades, with the systems of interest ranging from asymmetries in two-nucleon systems, through transitions in nuclei with $A\approx 20$, up to anapole moments of heavy nuclei, see, e.g., Refs.~\cite{Adelberger:1985ik,Haeberli:1995uz,RamseyMusolf:2006dz} and references therein.
It is not clear however, what conclusions can be drawn from these analyses. Not only is the DDH approach based on a number of assumptions about the short-distance details of the interactions, but the many-body physics involved in many of the observables considerably complicates the interpretation in terms of nucleon-nucleon interactions.

To avoid these difficulties, a systematic study of hadronic parity violation in few-nucleon systems based on effective field theory (EFT) was proposed in Ref.~\cite{Zhu:2004vw}, which contains a comprehensive analysis of PV interactions for EFTs with and without pions as dynamical degrees of freedom. 
For earlier EFT applications to hadronic parity violation see Refs.~\cite{Savage:1998rx,Kaplan:1998xi,Savage:1999cm,Savage:2000iv}. 
The EFT approach avoids model assumptions about the form of short-distance physics, and provides a consistent treatment of two-, three- and few-nucleon interactions as well as the coupling to external currents. In addition, theoretical errors can be estimated based on the the so-called power counting of the EFT.

\section{Parity-violating Lagrangian in \eftnopi}

Pionless EFT (\eftnopi) has been successfully applied to PC processes involving two and more nucleons, see, e.g., Refs.~\cite{Bedaque:2002mn,Platter:2009gz} for reviews.
The only dynamical degrees of freedom are nucleons, all other degrees of freedom including pions are integrated out.
The Lagrangian consists of all nucleon contact operators that are consistent with the symmetries of the underlying theory.
The short-distance details of the underlying theory are encoded in the so-called low-energy couplings (LECs) which accompany the operators in the EFT Lagrangian.
These LECs cannot be determined within the EFT.
While, in principle, they can be calculated in the underlying theory, in practice they are more often determined by comparison with experiment.
Once a LEC has been extracted, it can be used in the calculation of other observables.

For the case of low-energy PV interactions between two nucleons, the Lagrangian can be written in terms of five S-P wave transitions. 
In the following, the Lagrangian using dibaryon fields is used. 
Other forms of the Lagrangian can be found in Refs.~\cite{Zhu:2004vw,Girlanda:2008ts, Phillips:2008hn}.
At leading order (LO) the Lagrangian is given by \cite{Schindler:2009wd}
\begin{align}
  \label{eq:PVLag}
  \mathcal{L}_{PV}^d =& - \left[ \gA d_t^{i\dagger} \left(N^T
      \sigma_2 \tau_2\,i\LRd_i N\right) \right. \notag\\
  &\quad\quad +\gB d_s^{a\dagger}
  \left(N^T\sigma_2 \ \VS \cdot \tau_2 \tau_a \,i\LRd  N\right) \notag\\
  &\quad\quad +\gC \ \epsilon^{3ab} \, d_s^{a\dagger}
  \left(N^T \sigma_2  \ \VS\cdot \tau_2 \tau^b \LRd N\right) \notag\\
  &\quad\quad +\gD \ \mathcal{I}^{ab} \, d_s^{a\dagger}
  \left(N^T \sigma_2 \ \VS\cdot \tau_2 \tau^b \,i \LRd N\right) \notag\\
  &\quad\quad \left. +\gE \ \epsilon^{ijk} \, d_t^{i\dagger} \left(N^T \sigma_2
      \sigma^k \tau_2 \tau_3 \LRd{}^{j} N\right) \right] +\mathrm{h.c.}
  +\ldots,
\end{align}
where $d_t$ $(d_s)$ denotes the dibaryon field in the \threeS (\oneS) channel, the $\sigma_i$ ($\tau_a$) are spin (isospin) Pauli matrices, $a\, \mathcal{O}\LRd b = a\,\mathcal{O}\vec D b - (\vec D a)\mathcal{O} b\
$ with $\mathcal{O}$ some spin-isospin-operator, and
\begin{equation}
\mathcal{I}=
\begin{pmatrix}
1 & 0 & 0 \\
0 & 1 & 0\\
0 & 0 & -2
\end{pmatrix}. 
\end{equation}
Determination of the PV LECs $g$ requires measurements of five independent PV observables.
Given the difficulty of the corresponding experiments, this presents a considerable challenge. 
It also requires the calculation of at least five independent PV observables in a consistent formalism.
In the following the first results of a concentrated effort towards this goal are presented. 
All results are adjusted to the dibaryon formalism of Eq.~\eqref{eq:PVLag} in combination with the conventions of Ref.~\cite{Griesshammer:2011md}.

\section{Two-nucleon sector}

The longitudinal asymmetry in $\vec{N}N$ scattering is defined as
\begin{equation}
A_L = \frac{\sigma_+ - \sigma_-}{\sigma_+ + \sigma_-}\, ,
\end{equation}
where $\sigma_\pm$ is the total cross section for scattering of a beam with $\pm$ helicity. 
The LO calculation of Ref.~\cite{Phillips:2008hn} finds (adjusted to the dibaryon conventions of Eq.~\eqref{eq:PVLag})
\begin{align}
A_L^{nn} & = - \sqrt{\frac{32 M}{\pi}}\,p \left( \gB-\gC+\gD \right)\,, \\
A_L^{pp} & = - \sqrt{\frac{32 M}{\pi}}\,p \left( \gB+\gC+\gD \right)\,, \\
A_L^{np} & = - \sqrt{\frac{32 M}{\pi}}\,p \frac{ \frac{d\sigma^{\oneS}}{d\Omega}}{\frac{d\sigma^{\oneS}}{d\Omega}+3\frac{d\sigma^{\threeS}}{d\Omega}}\left( \gB-2\gD \right)\,, \notag \\
& \quad - \sqrt{\frac{32 M}{\pi}}\,p \frac{ \frac{d\sigma^{\threeS}}{d\Omega}}{\frac{d\sigma^{\oneS}}{d\Omega}+3\frac{d\sigma^{\threeS}}{d\Omega}}\left( \gA+2\gE \right)\,,
\end{align}
where Coulomb corrections are neglected in the $\vec{p}p$ case. 
As shown in Ref.~\cite{Phillips:2008hn}, including Coulomb interactions amounts to corrections of $\approx 3\%$ at the lowest energy considered experimentally \cite{Eversheim:1991tg}, much smaller than the expected errors of a LO calculation.

The PV $np$ forward scattering amplitude can also be related to the spin rotation angle of a transversely polarized neutron beam traversing a hydrogen target. 
The next-to-leading-order (NLO) result of Ref.~\cite{Griesshammer:2011md} for the rotation angle per unit length is
\begin{align}
\frac{1}{N} \, \frac{d\phi_{PV}^{np}}{dl} = 4\sqrt{2\pi M} \left( \frac{\gA+2\gE}{\gamma_t} \frac{Z_t+1}{2} + \frac{\gB-2\gD}{\gamma_s} \frac{Z_s+1}{2} \right),
\end{align}
where $N$ is target density, $Z_{s/t} = \frac{1}{1-\gamma_{s/t}\rho_{s/t}}$, $\rho_{s/t}$ are the effective ranges in the spin-singlet/spin-triplet channel, and $\gamma_{s/t}$ are the poles in the $NN$ scattering amplitudes in the respective spin channels.

By considering different polarizations, two independent PV observables can be extracted from the reactions $np\leftrightarrow d\gamma$. 
The angular asymmetry $A_\gamma$ in the capture of polarized neutrons ($\vec{n}p\to d\gamma$) has been of particular interest, as in the meson-exchange picture it is dominated by one-pion exchange. 
In this picture, determination of $A_\gamma$ would provide constraints on the PV pion-nucleon coupling. 
However, at the very low energy at which the corresponding experiment is performed treating the pion as a dynamical degree of freedom is not necessary, and $A_\gamma$ constrains the LEC $\gE$ of Eq.~\eqref{eq:PVLag}.
At LO, the observable is given by \cite{Schindler:2009wd}
\begin{equation}
A_\gamma = \frac{4}{3} \sqrt{\frac{2}{\pi}} \frac{M^\frac{3}{2}}{\kappa_1(1-\gamma_t a^{\oneS})}\,\gE\,,
\end{equation}
where $\kappa_1$ is the nucleon isovector anomalous magnetic moment and $a^{\oneS}$ the scattering length in the spin-singlet channel.
The angular asymmetry was also calculated in pionless EFT including effective-range corrections in Ref.~\cite{Savage:2000iv}.
An ongoing experimental effort at Oak Ridge's Spallation Neutron Source is devoted to measuring $A_\gamma$ \cite{Gericke:2011zz}.

The PV interactions induce a circular photon polarization $P_\gamma$ in the capture of unpolarized neutrons ($np\to d\vec{\gamma}$). 
This observable is independent of $A_\gamma$ and provides complementary information on the LECs.
At LO in \eftnopi it is given by \cite{Schindler:2009wd}
\begin{equation}
P_\gamma = -2\sqrt{\frac{2}{\pi}} \frac{M^\frac{3}{2}}{\kappa_1(1-\gamma_t a^{\oneS})}\left[ \left( 1-\frac{2}{3}\gamma_t a^{\oneS} \right)\gA +\frac{\gamma_t a^{\oneS}}{3} \left( \gB-2\gD \right) \right].
\end{equation}
This polarization was also determined in Ref.~\cite{Shin:2009hi}.
For exactly reversed kinematics, $P_\gamma$ is identical to the asymmetry $A_L^\gamma$ in the break-up reaction $\vec{\gamma}d\to np$. A measurement of $A_L^\gamma$ might be feasible at a proposed upgrade of the HIGS facility at the Triangle Universities Nuclear Laboratory.

\section{Three-nucleon sector}

The power counting of \eftnopi in principle predicts at which order three-nucleon (3N) interactions appear.
In the PC sector, naive application of the power counting indicates that 3N interactions first contribute at next-to-next-to-leading order (NNLO). However, as shown in Refs.~\cite{Bedaque:1998kg,Bedaque:1999ve}, a 3N interaction is needed already at LO to remove cutoff dependence in the $nd$ scattering amplitude in the spin-doublet channel. Parity-violating 3N interactions are similarly predicted to appear at  NNLO, but the ``promotion'' of the corresponding PC terms makes a more detailed study of these operators necessary. Such an analysis was performed in Ref.~\cite{Griesshammer:2010nd}, which showed that no PV 3N interactions are required at LO and NLO. This ensures that, up to an accuracy of $\approx 10\%$, also PV three-nucleon observables can be analyzed in terms of PV 2N interactions alone. Given the difficulty of extracting even these 5 LECs, the presence of additional 3N interactions with accompanying unknown LECs would have complicated the comprehensive analysis significantly.

Parity violation in neutron-deuteron scattering can be studied in $nd$ spin rotation. Reference \cite{Griesshammer:2011md} finds that the rotation angle up to NLO in \eftnopi is given by
\begin{align}
\frac{1}{N} \, \frac{d\phi_{PV}^{nd}}{dl} = \left\{ [16 \pm 1.6]\gA + [34\pm 3.4]\gE + [4.6 \pm 1.0] \left(3\gB - 2\gD\right) \right\}.
\end{align}
This calculation also confirms numerically that a PV 3N interaction is not required for the renormalization of the PV scattering amplitude. A LO result of the spin rotation angle can also be found in Ref.~\cite{Vanasse:2011nd}, which in addition contains results for longitudinal asymmetries in $nd$ scattering.

\section{Conclusions and outlook}

The presented results form part of a comprehensive study of hadronic parity violation in few-nucleon systems.
The use of \eftnopi ensures that PC and PV interactions as well as external currents are treated in a unified framework.
Theoretical errors can be estimated based on power counting, which also provides the possibility of systematic improvements of the present calculations.
Up to NLO, observables can be parameterized in terms of five PV LECs corresponding to the five LO S-P wave operators in two-nucleon interactions, while PV three-nucleon interactions do not start to contribute until at least NNLO.
The five LO LECs are currently not determined. However, upcoming experimental results promise to provide crucial information. 
In addition, the first lattice QCD study of a PV pion-nucleon coupling was recently performed \cite{Wasem:2011tp}.
The determination of LECs from lattice QCD provides a promising alternative and/or cross check to the extraction from experimental results.
\eftnopi calculations of further three- and few-nucleon observables will provide additional constraints and will be important in improving our understanding of hadronic parity violation at low energies.

\acknowledgments

I thank H.~W.~Grie{\ss}hammer, D.~R.~Phillips, and R.~P.~Springer for their collaboration on the calculations described here as well as for many interesting and stimulating discussions.

\end{document}